\documentclass[11pt]{article}
\usepackage[utf8]{inputenc}
\usepackage{graphicx}
\usepackage[utf8]{inputenc}
\usepackage{amsmath}
\usepackage{amsfonts}
\usepackage{amssymb}
\usepackage{graphicx}
\usepackage{setspace}
\usepackage{authblk}
\usepackage{caption}
\usepackage{float}
\usepackage{cite}
\usepackage[left=2cm,right=2cm,top=1cm,bottom=2cm]{geometry}
\usepackage{color}
\usepackage{color,soul}
\title{\textbf{Buchdahl Spacetime with Compact Body Solution of Charged Fluid and Scalar Field Theory}}

\author[1]{Manuel Malaver de la Fuente* and Rajan Iyer**\par\small{Bijective Physics Institute, Idrija, Slovenia*\par Maritime University of the Caribbean,  Department of Basic Sciences, Catia la Mar, Venezuela.*\par Environmental Materials Theoretical Physicist, Department of Physical Mathematics Sciences Engineering   Project Technologies, Engineeringinc International Operational Teknet Earth Global, Tempe, Arizona, United States of America**\par Email: mmf.umc@gmail.com*\par Email: engginc@msn.com**}\\

\par\hspace{1.0cm}
\large{Alokananda Kar' and Shouvik Sadhukhan''}\par\small{Department of Physics; University of Calcutta; 92 APC Road, Kolkata 700009, West Bengal, India'\par Department of Physics; Indian Institute of Technology (ISM), Dhanbad; Police Line Road, Main Campus IIT (ISM, near Rani Bandh, Hirapur, Sardar Patel Nagar, Dhanbad, Jharkhand 826004, India' \par Department of Physics; Indian Institute of Technology, Kharagpur 721302, West Bengal, India''\par Department of Physics; Indian Institute of Space Science and Technology, Thiruvananthapuram; Valiamala Road, Valiamala, Kerala 695547''\par Email: alokanandakar@gmail.com'\par Email: shouvikphysics1996@gmail.com''}\\

\par\hspace{1.5cm}
\par
\large{Sudhaker Upadhyay''' and Ertan Güdekli*"}\par\small{Department of Physics, Magadh University, Bodh Gaya, Bihar 824234, India'''\par Email: sudhakerupadhyay@gmail.com \par Department of Physics, Istanbul University, 34452 Istanbul, Turkey*"  \par Email: gudekli@istanbul.edu.tr*"}}

\date{Dated : \today}
\begin{document}
\maketitle

\begin{abstract}
The present work contains a discussion on compact stellar body that is influenced by the effect of electromagnetic fields. We have tried to discuss a solution of Einstein-Maxwell Field equation with the interpretation of Buchdahl space time type function. The interior fluid density and anisotropic pressures (tangential pressure $p_t$ and radial pressure $p_r$) have been derived for a charged fluid of compact body which is also continuous with the Reissner–Nordstrom metric exterior solution. We have discussed the singularity free interior anisotropic fluid under electromagnetic influence. The thermodynamics energy conditions and their variation with interior radius have been discussed here. We have also discussed the physical acceptability of this new model. Finally we have included the scalar field theory corresponding to the interior fluid that helped us to discuss the compact body strange star evolution with time. The generalized entropy (Bekenstein-Hawking entropy) evolution of the compact body has also been introduced in the discussion of interior fluid evolution analysis.\\

\end{abstract}

\section{Introduction}
The structure and life span of compact bodies or strange stars are the biggest problems in astrophysics. The astrophysical bodies like nebula, galaxies, stars  mainly follow the local solutions of Einstein equation in General Theory of Relativity. Hence the structure and their interior fluid stability became the biggest mystery after the application of Einstein equation in fluid mechanics and Astrophysics. The astronomical bodies mainly face two types of forces viz. gravitational force due to its field and Hydrodynamical force. The equilibrium condition between those forces may provide the mass versus radius limits on the astrophysical systems like stars. The hydrodynamical pressures from the ordinary fluids and its limits do not match the observations and hence the idea of interpreting different kind of fluid models has been brought in astrophysics. Different kind of fluid models, Dark energy models, Field models have been introduced here to discuss the interior fluid mechanism of stars and astrophysical bodies. perfect fluid models, polytropic fluid models, Chaplygin gas models, Inhomogeneous and anisotropic models and Nonlinear fluid models are some of the examples. The variation of fluid models can provide the variation of the equilibrium conditions and hence we can find different limits of the Mass-Radius ratios of the astrophysical bodies.\cite{1,2,3,4,5,6,7,8,9,10,11,12,13,14,15,16,17,18,19,20,21,22,23,24,25,26,27,28,29,30}\par

Recent observation provides the existence of a compact star with mass $2.59^{+0.08}_{-0.09}M_{\odot}$ (Abbott et al. (2020)). From the references of (Bailyn et al. 1998; Ozel et al. 2010; Belczynski et al. 2012) it can be observed that this mass remains within the mass gap of stellar bodies. Thus the ambiguity comes in the discussion of interior fluid of those compact bodies. From literatures of theoretical and Observational studies it can be found that the blackholes generated from stellar evolution must have the masses more than $5M_{\odot}$ whereas the stable neutron star should have a mass of mostly within $3M_{\odot}$. The neutron star with mass more than that can loose its individuality hypothetically. This kind of problems have been studied with the introduction of quark stars (The intermediate phase of starry evolution between neutron star and black hole). Different kinds of quark fluid function have been assumed to reconstruct the interior fluid model of a star or compact body. The estimated radius of some objects like  (LMC X-4, 4U 1820-30, Her X-1, etc.) suggest that their characteristics and phenomenology are similar to those of quark stars. On the other hand the gravitational wave signal GW170817 don't allow the large mass of that compact object instead suggests around $2.5M_{\odot}$. The heaviest neutron star possible is around $2.01\pm0.04M_{\odot}$. Hence that large values of masses of compact objects within $M\in(2.5,5)M_{\odot}$ is theoretically impossible.\cite{55,56,57,58,59,60,61,62,63,64,65,66,67,68,69,70}\par

The gamma ray and X-ray blasts are the results of the electron-proton interactions with the molecular clouds. The electrons and protons are accelerated from the supernova explosion shock waves or pulsar wind nebulae. This supernova  generally happens from the non-equilibrium conditions found inside of a compact body. The hydrodynamical balance destruction is the primary cause behind those kinds of blasts. Hence, Compact body interior fluid hydrodynamics study is important in astrophysics. Recent observational an computational analysis (Kar et.al (2021), Abeysekara, A. U et.al (2020), Cao et al. (2021), Mondal et al (2021)) provides several examples of such high energy explosions. Hence the idea of modifications of interior fluid dynamics and the equilibrium conditions caused by those hydrodynamical and gravity pressures and their interactions became a necessity in astrophysics.\cite{40,71,72,73,74,75,76,77,78,79,80}\par

The interior structure reconstruction of a compact body like stars, pulsars, nebulae, giant molecular clouds etc. can be followed by several types of fluid models. Different fluid models provide different equation of states that again reconstruct the nature of fluid pressures and hydrodynamic properties. Several dark energy models like fluid dark energies, holographic dark energies can be used in this purpose. From cosmological point of view the dark energy models have extensive use in discussion of the origin of negative pressure that causes the accelerated expansion of universe. If we use this kind of fluids in the discussion of astrophysical bodies interior fluid mechanism, we can modify the equilibrium conditions of them. In gravitational point of view these compact bodies generally follow the local solutions of Einstein equation which is a result of generalized action. The action can be coupled with different kinds of fluids to find some new type interior solutions. Recent observations and theoretical analysis proved that the local fluids may have anisotropy and Inhomogeneity under certain conditions. The inhomogeneities mainly brought with the fluid viscosity, fluid-fluid interactions, Fluid-fields interactions. There are several literatures available on anisotropic fluid solutions (Malaver et.al (2013-2021)\cite{42,43,44,45,46,47,48,49,50,51,52,53,54}), Inhomogeneous fluid solutions (Kar et.al (2021) and Sadhukhan et.al (2020) \cite{31,32,33,34,35,36,37,38,39}). The nonlinear fluid models are the one that can represent the hydrodynamical properties with non linear equation of states which also a different representations of dark energy (Kar et.al (2021) and Sadhukhan et.al (2020) \cite{35,37}). The scalar field theories can also represent the hydrodynamic nature of the interior fluid systems in scalar field potential variation approach. In general the scalar field theory generalizes the field interaction models inside of a compact body that can again modify the equilibrium conditions of the same. The scalar field models are mainly used to discuss the expanding nature of universe which  can also discuss the mass accretion of any compact body. This mass accretion helps in the evolution of these compact bodies and unifies the cosmological expanding models with astrophysical compact body interior solutions. Hence with scalar field models we can discuss the evolution of compact body interior properties. The thermodynamics energy conditions in such problems are mainly assumed to be the stability conditions of the models. Although the violation of any of these conditions may bring some shift in equilibrium of compact bodies (Kar et.al (2021) \cite{34,35} and Malaver et.al (2021)\cite{39}). The interior fluid interactions and few other conditions may bring local anisotropy in the interior fluid systems. One of the most successful solution of such anisotropic system is Buchdahl space time solutions ( Buchdahl et. al (1959)). Several applications have been done on it till date (Malaver et.al (2013 to 2021)\cite{39,42,43,44,45,46,47,48,49,50,51,52,53,54} and Kumar et.al (2021)). \par

In this paper we have mainly discussed the nature of interior fluid of a compact object with new kind of solutions of Buchdahl type functions. We have introduced the anisotropic solutions and perfect fluid model as interior fluid. The physical acceptability conditions, junction conditions, stability conditions and thermodynamical energy conditions have been discussion with the new model of compact star. The corresponding scalar field and its potential have been found that can help to interpret the inside interaction pictures. The equilibrium conditions with different force components have been discussed with this model. Finally, we have introduced the generalized entropy of this system and its time evolution with the pre-assumed scale factor solution of isotropic-homogeneous universe model (Friedmann model). The mass accretion of that compact object and its relation with the interior properties have been discussed here.\par

The manuscript is prepared as follows. In section 2 we have given the mathematical basis of compact stars. In section 3 we have discussed the acceptability and energy conditions of our newly derived model. In section 4 we have discussed the interior thermodynamics of our compact body. In section 5 we have given a scalar field theory for our model. Section 6 contains the interpretations of mass change and time dependency in scalar field theory. We conclude the paper with results analysis and discussion in section 7 and 8.

\section{Mathematical Basis and Methodology of Compact Stars Astrophysics}
The work should start from the electromagnetism coupled gravity action which can be written as follows.

\begin{equation}
    S=\int{d^4x\sqrt{-g}(f(R,T)+L_m+L_e)}
\end{equation}
We can apply the least action principle on the above action to get equation of motion. Here, $L_e=$ Lagrangian for electromagnetic fields and $L_m=$ the Lagrangian for matter fields. The following equation is the outcome of the variation principle from action in equation 1.

\begin{equation}
    \frac{\partial f(R,T)}{\partial R}R_{\mu\nu}-\frac{1}{2}g_{\mu\nu}f(R,T)+(g_{\mu\nu}\Box-\triangledown_{\mu} \triangledown_{\nu})\frac{\partial f(R,T)}{\partial R}=8\pi(T_{\mu\nu}+E_{\mu\nu})-\frac{\partial f(R,T)}{\partial T}(T_{\mu\nu}+\Theta_{\mu\nu})
\end{equation}
where
\begin{equation}
    \Theta_{\mu\nu}=g^{\alpha\beta}\frac{\partial T_{\alpha\beta}}{\partial g^{\mu\nu}}
\end{equation}
And ,
\begin{equation}
    \Box=\frac{1}{\sqrt{-g}}\partial_{\mu}(\sqrt{-g}g^{\mu\nu}\partial_{\nu})
\end{equation}
The electromagnetic Lagrangian $L_e$ can produce the EM component of the action and the energy tensor on EM field can be written as following equation.
\begin{equation}
    E_{\mu\nu}=\frac{1}{4\pi}(F^{\alpha}_{\mu}F_{\nu\alpha}-\frac{1}{4}F^{\alpha\beta}F_{\alpha\beta}g_{\mu\nu})
\end{equation}
Now, the energy momentum tensor can be written also as follows.
\begin{equation}
    T_{\mu\nu}=g_{\mu\nu}L_m-2\frac{\partial L_m}{\partial g_{\mu\nu}}
\end{equation}

\subsection{Einstein-Maxwell Field Equations with Spherical Symmetric, Homogeneous and Isotropic Frame}
Here we produced the field equations that are modified form of Einstein equation with electromagnetic fields. Now using $f(R,T)=R$ we can get the Einstein field equation as follows.
\begin{equation}
    G_{\mu\nu}=8\pi(T_{\mu\nu}+E_{\mu\nu})=8\pi T_{\mu\nu}^{eff}
\end{equation}
Where,
\begin{equation}
    T_{\mu\nu}^{eff}=diag(-\rho-\frac{1}{2}E_{f}^2,p_r-\frac{1}{2}E_{f}^2,p_t+\frac{1}{2}E_{f}^2,p_t+\frac{1}{2}E_{f}^2)
\end{equation}
The primary aim of this work is to consider the anisotropic geometry reconstruction using Buchdahl space and hence we have chosen the generalised line element as follows.
\begin{equation}
    ds^2=-e^{2\nu(r)}dt^2+e^{2\lambda(r)}dr^2+r^2d\Omega^2
\end{equation}
With the above line element we can obtain the following Einstein-Maxwell field equations.
\begin{equation}
    \frac{1}{r^2}(1-e^{-2\lambda})+\frac{2\lambda'}{r}e^{-2\lambda}=\rho+\frac{1}{2}E_{f}^2
\end{equation}
\begin{equation}
    -\frac{1}{r^2}(1-e^{-2\lambda})+\frac{2\nu'}{r}e^{-2\lambda}=p_r-\frac{1}{2}E_{f}^2
\end{equation}
\begin{equation}
    e^{-2\lambda}(\nu''+\nu'^2+\frac{\nu'}{r}-\nu'\lambda'-\frac{\lambda'}{r})=p_t+\frac{1}{2}E_{f}^2
\end{equation}
\begin{equation}
    \sigma=\frac{1}{r^2}e^{-\lambda}(r^2E_{f})'
\end{equation}
All the above equations follow the kinematics of anisotropic matter system. These equations help us to discuss the compact body model with Buchdahl solution assumptions. Here we have considered $\frac{8\pi G}{c^2}=1$ and $c=1$.

\subsection{Buchdahl Spacetime geometry with new function}
Now we apply the Buchdahl space-time in equations 10 to 13 to solve and find our model of compact bodies. We assume the following transformations.
\begin{center}
    $x=cr^2$ ; $z(x)=e^{-2\lambda(r)}$ and $A1^2y^2(x)=e^{2\nu(r)}$
\end{center}
Hence the line element will become as follows.
\begin{equation}
     ds^2=-A1^2y^2dt^2+\frac{1}{4cxz}dx^2+\frac{x}{c}d\Omega^2
\end{equation}
The equations of kinematics from 10 to 13 can be re-written as follows.
\begin{equation}
    \frac{1-z}{x}-2\dot{z}=\frac{\rho}{c}+\frac{E_{f}^2}{2c}
\end{equation}
\begin{equation}
    4z\frac{\dot{y}}{y}+\frac{z-1}{x}=\frac{p_r}{c}-\frac{E_{f}^2}{2c}
\end{equation}
\begin{equation}
    4xz\frac{\ddot{y}}{y}+(4z+2x\dot{z})\frac{\dot{y}}{y}+\dot{z}=\frac{p_t}{c}+\frac{E_{f}^2}{2c}
\end{equation}
\begin{equation}
    \frac{\sigma^2}{c}=\frac{4z}{z}(x\dot{E_{f}}+E_{f})^2
\end{equation}
Additionally we must follow the fluid nature of the inner compact bodies. Here we have used the perfect fluid. We considered anisotropic fluid that provides two pressures on the system. Hence we can write as follows.
\begin{equation}
    p_r=\omega\rho
\end{equation}
And,
\begin{equation}
    p_t=p_r+\Delta
\end{equation}
Therefore including all the equations from 15 to 20 we can have six equations with eight unknowns $(\rho, p_r, p_t, \Delta, E_{f}, \sigma, y, z)$. Hence, to solve these equations we choose two of the unknowns. These choices are represented as Buchdahl geometry.
\begin{equation}
    z(x)=\frac{k+x}{k(1+x)}
\end{equation}
And the electromagnetic field energy as follows.
\begin{equation}
    \frac{E_{f}^2}{2c}=x(a+bx)
\end{equation}
With these two assumptions we now have six unknowns which can be solved using the six equations. Here the term $z(x)$ is the gravitational potential that depends on the geometry of the compact object. The metric potential is regular and continuous at the origin and well-behaved in the interior of the sphere. The electromagnetic field energy tensor is finite at the centre of the star and continuous as well. Along with all those variables another variable is much important in the discussion of compact bodies. This is the spatial dependent mass of of the compact star. This can be defined as follows.
\begin{equation}
    M(x)=\frac{1}{4c^{3/2}}\int_{0}^{x}{\sqrt{x}(\rho+E_{f}^2)dx}
\end{equation}

\subsection{Solution of Einstein-Maxwell equation}
Now we solve the variables using the Friedmann type kinematical equations of the compact body. From equation 15 using our assumptions we can write the energy density of the fluid system inside the compact body as follows.
\begin{equation}
    \rho=\frac{c((k-1)(x+3)-kx(a+bx)(x+1)^2)}{k(x+1)^2}
\end{equation}
Now from equation 19 we can find the radial component of pressure using this energy density.
\begin{equation}
    p_r=\frac{\omega c((k-1)(x+3)-kx(a+bx)(x+1)^2)}{k(x+1)^2}
\end{equation}
Now from equation 16 using the definition of $p_r$ and $E$ as well as $z(x)$, we may write,
\begin{equation}
    y(x)=c_1(x+1)^A(k+x)^B\exp{(\frac{(1+\omega)}{24}(Cx^3+Dx^2+Ex))}
\end{equation}
Now the mass function can be defined as follows by following equation 23.
\begin{equation}
    M(x)=\frac{x^{3/2}(-5kbx^3-(7a+5b)kx^2-7kax+35k-35)}{70k\sqrt{c}(x+1)}
\end{equation}
This function provides the variation of mass of the compact body with radial direction. Now from equation 18 we can derive the radial dependent functional form of of shearing scalar.
\begin{equation}
    \sigma^2=\frac{2c^2(k+x)(4bx+a)^2}{k(x+1)(a+bx)}
\end{equation}
The tangential pressure can be derived from the equations 17, 21 and 22. The radius dependent functional form can be written as follows.
\begin{equation}
\begin{split}
    p_t=\frac{\omega c((k-1)(x+3)-kx(a+bx)(x+1)^2)}{k(x+1)^2}\\+\frac{4xc(k+x)}{k(1+x)}(\frac{A^2-A}{(1+x)^2}+\frac{2AB}{(x+1)(k+x)}+\frac{(1+\omega)(3Cx^2+2Dx+E)}{12(x+1)}\\+\frac{B^2-B}{(k+x)^2}+\frac{B(1+\omega)(3Cx^2+2Dx+E)}{12(k+x)}+\frac{(\omega+1)(6Cx+2D)}{24}+\frac{(\omega+1)^2(3Cx^2+2Dx+E)}{576})\\+\frac{(1-k)c}{k(1+x)^2}(1+2x(\frac{A}{1+x}+\frac{B}{k+x}+\frac{(1+\omega)(3Cx^2+2Dx+E)}{24}))+\frac{c(k-1)}{k(1+x)}-2xc(a+bx)
\end{split}
\end{equation}
Now the anisotropy in pressure can be defined as follows.
\begin{equation}
\begin{split}
    \Delta=\frac{4xc(k+x)}{k(1+x)}(\frac{A^2-A}{(1+x)^2}+\frac{2AB}{(x+1)(k+x)}+\frac{(1+\omega)(3Cx^2+2Dx+E)}{12(x+1)}\\+\frac{B^2-B}{(k+x)^2}+\frac{B(1+\omega)(3Cx^2+2Dx+E)}{12(k+x)}+\frac{(\omega+1)(6Cx+2D)}{24}+\frac{(\omega+1)^2(3Cx^2+2Dx+E)}{576})\\+\frac{(1-k)c}{k(1+x)^2}(1+2x(\frac{A}{1+x}+\frac{B}{k+x}+\frac{(1+\omega)(3Cx^2+2Dx+E)}{24}))+\frac{c(k-1)}{k(1+x)}-2xc(a+bx)
\end{split}
\end{equation}
Here we have used five constants which can be written as follows.
\begin{center}
    $A=\frac{1}{2}\omega$ ;  $B=\frac{(1+\omega)(bk^4-(a+b)k^3+ak^2+k)-(1+3\omega)}{4}$
\end{center}
\begin{center}
    $C=-2bk$ ; $D=k(bk-(a+b))$ ; $E=6k(-bk^2+(a+b)k-a)$
\end{center}

\begin{figure}[H]
\centering
\begin{minipage}[b]{0.4\textwidth}
    \includegraphics[width=\textwidth]{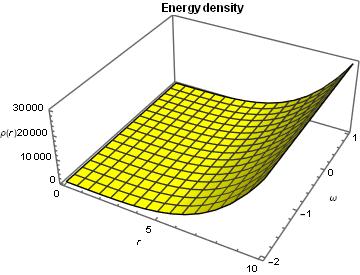}
    \caption{Graph for Energy Density with $a=2$, $b=-3$, $c=1$ and $k--5$}
\end{minipage}
\hfill
\begin{minipage}[b]{0.4\textwidth}
    \includegraphics[width=\textwidth]{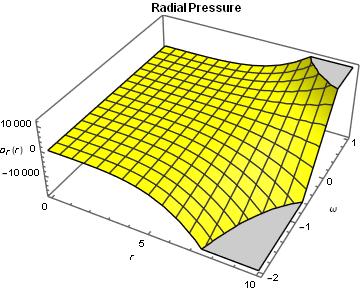}
    \caption{Graph for Radial Pressure with $a=2$, $b=-3$, $c=1$ and $k--5$}
\end{minipage}
\end{figure}
\begin{figure}[H]
\centering
\begin{minipage}[b]{0.4\textwidth}
    \includegraphics[width=\textwidth]{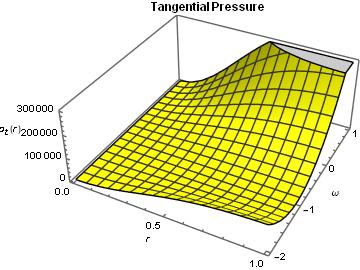}
    \caption{Graph for Tangential Pressure with $a=2$, $b=-3$, $c=1$ and $k--5$}
\end{minipage}
\hfill
\begin{minipage}[b]{0.4\textwidth}
    \includegraphics[width=\textwidth]{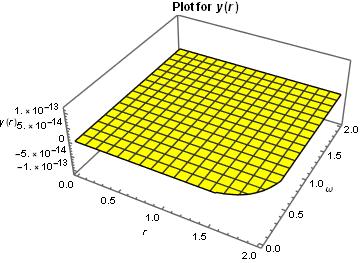}
    \caption{Graph for $Y(r)$ with $a=2$, $b=-3$, $c=1$ and $k--5$}
\end{minipage}
\end{figure}

\begin{figure}[H]
\centering
\begin{minipage}[b]{0.4\textwidth}
    \includegraphics[width=\textwidth]{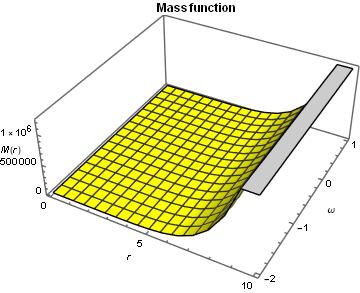}
    \caption{Graph for Mass Function with $a=2$, $b=-3$, $c=1$ and $k--5$}
\end{minipage}
\hfill
\begin{minipage}[b]{0.4\textwidth}
    \includegraphics[width=\textwidth]{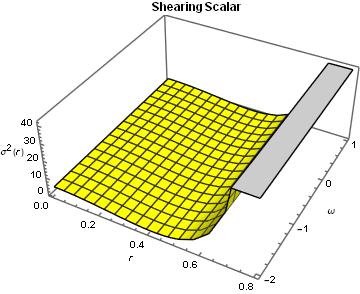}
    \caption{Graph for Shearing Scalar with $a=2$, $b=-3$, $c=1$ and $k--5$}
\end{minipage}
\end{figure}
\begin{figure}[H]
\centering
\begin{minipage}[b]{0.4\textwidth}
    \includegraphics[width=\textwidth]{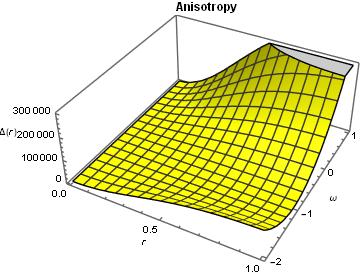}
    \caption{Graph for Anisotropy with $a=2$, $b=-3$, $c=1$ and $k--5$}
\end{minipage}
\end{figure}

\subsection{Comparison between our new model solution and Reissner-Nordstrom metric}
The solution discussed above represents a charged compact object whose geometry is modified with the electromagnetic action and the interior fluid system follows the perfect fluid mechanism. Hence we can assume the exterior portion of that object satisfies the Reissner-Nordstrom solution of field equation. thus, considering the total mass of the system equal to $M$, we can write the following line element solution of Einstein-Maxwell field equation.
\begin{equation}
    ds^2=-(1-\frac{2M}{r}+\frac{Q^2}{r^2})dt^2+\frac{dr^2}{(1-\frac{2M}{r}+\frac{Q^2}{r^2})}+r^2d\Omega^2
\end{equation}
Here $r$ is the radial distance of any exterior point from the center of the body and $Q$ is the total charge of the same.

\section{Physical Analysis and Acceptability conditions of newly derived model}
Here we discuss the validity of newly derived model. The acceptability conditions provide the physical viability and stability of our model to discuss the compact bodies. The acceptability conditions are mainly contained with the following conditions and points.
\begin{itemize}
    \item Matching Conditions
    \item Compactness
    \item Regularity and Maximality criterion
    \item Thermodynamics Energy Conditions
    \item $\rho \geq 0$ for $0\geq R \geq a$ where $R$ is the total radius
    \item $p_r\geq 0$ and $p_t\geq 0$ for $0\geq R \geq a$
    \item $p_r=0$ at $r=0$
    \item $\frac{dp_r}{d\rho}=v_{r}^2\geq 0 < c$ and $\frac{dp_t}{d\rho}=v_{t}^2\geq 0 < c$ 
    \item Adiabatic Index $\gamma=\frac{(c^2\rho+p_r)}{p_r}\frac{dp_r}{c^2d\rho}=v_{r}^2\geq 1$ or $4/3$ and $\gamma=\frac{(c^2\rho+p_t)}{p_t}\frac{dp_t}{c^2d\rho}=v_{t}^2\geq 1$ or $4/3$
    \item Stability under forces
\end{itemize}
We proceed with our previously derived energy density and pressure of the compact body interiors.
\subsection{Matching Conditions}
Matching conditions are the boundary conditions which match the exterior and interior compact body solutions of Einstein equation at the boundary or on the surface of the system i.e. at $r=a$. Hence the conditions are discussed as follows.
\subsubsection*{Junction Condition 1 : $g_{\mu\nu}(a+0)=g_{\mu\nu}(a-0)$}
here we must have to satisfy the following relation. We have used the radius of the compact bodies as $R$. (from Equations 24 to 31)
\begin{equation}
    1-\frac{2M}{R}+\frac{Q^2}{R^2}=A1^2y^2(cR^2)
\end{equation}
And,
\begin{equation}
    1-\frac{2M}{R}+\frac{Q^2}{R^2}=z(cR^2)
\end{equation}
Hence using the second equation we can find the value of $k$ as follows.
\begin{equation}
    k=(\frac{(1+R)}{R}(1-\frac{2M}{R}+\frac{Q^2}{R^2}-\frac{1}{1+R}))^{-1}
\end{equation}

\subsubsection*{Junction Condition 2 : $\partial_{r}g_{\mu\nu}(a+0)=\partial_{r}g_{\mu\nu}(a-0)$}
Here we can discuss the metric change through the surface of the compact body and hence we use $g_{tt}$ in place of $g_{\mu\nu}$. Hence we get the following relations and results. (from Equations 24 to 31)
\begin{equation}
    \frac{M}{R^2}-\frac{Q^2}{R^3}=2A1^2y(cR^2)y'(cR^2)
\end{equation}
Hence using this relation and the equation 32 we get the values of constant $A$.
\begin{equation}
    A1=(\frac{1}{2y(cR^2)y'(cR^2)+\frac{y^2(cR^2)}{R}}(\frac{1}{R}-\frac{M}{R^2}))^{1/2}
\end{equation}

\subsubsection*{Junction Condition 3 : $p(a+0)=p(a-0)$}
Here using the definitions of pressures of the interior fluid system of the newly designed compact body, we may write following relation. (from Equations 24 to 31)
\begin{equation}
    \omega c((k-1)(R+3)-kR(a+Rb)(R+1)^2)
\end{equation}
As we know that the constant $omega\neq 0$, hence we get the following relation for $a$ and $b$.
\begin{equation}
    a+Rb=\frac{R+3}{R(R+1)^2}[1-(\frac{1+R}{R})(1-\frac{2M}{R}+\frac{Q^2}{R^2}-\frac{1}{1+R})]
\end{equation}

\subsection{Compactness and Surface Red shift}
Compactness is a parameter to identify any compact object and classify them among all the list of objects. The classification can be done with the values of the Compactness parameter and the functional form is as follows.
\begin{equation}
    U=\frac{M(R)}{R}
\end{equation}
The classification can be provided as follows.
\begin{itemize}
    \item $U\approx 10^{-5}$ for Normal Star
    \item $U\approx 10^{-}$ for White Dwarfs
    \item $10^{-1} < U < \frac{1}{4}$ for neutron Star
    \item $\frac{1}{4} < U < \frac{1}{2}$ for Ultra Compact Star
    \item $U\approx \frac{1}{2}$ for Blackholes
\end{itemize}
In our present work we have introduced Buchdahl function in the solution of Einstein-Maxwell filed equation locally. Hence, the defined values for this compact object should be beyond the list given above. Although, we have tried to give the values of compact object with radius equivalent to solar radius.

\subsection{Regularity and Maximally Criterion for Pressure and Density}
The regularity criterion is the condition to make any compact object stable in its structure. The stability happens with the variation of pressure and density of the interior fluids with radius starting from centre. The criterion can be established as follows.
\begin{itemize}
    \item $\frac{d\rho}{dr} < 0$
    \item $\frac{dp_r}{dr} < 0$
    \item $\frac{dp_t}{dr} < 0$
    \item $\frac{d\Delta}{dr} < 0$
\end{itemize}
The violation of those conditions can provide boundary blast and hence any kind of compact object solutions should become unstable.

\begin{figure}[H]
\centering
\begin{minipage}[b]{0.4\textwidth}
    \includegraphics[width=\textwidth]{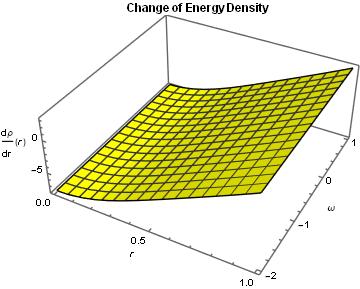}
    \caption{Graph for Change Energy Density with $a=2$, $b=-3$, $c=1$ and $k--5$}
\end{minipage}
\hfill
\begin{minipage}[b]{0.4\textwidth}
    \includegraphics[width=\textwidth]{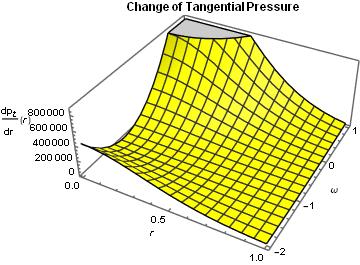}
    \caption{Graph for Change of Tangential Pressure with $a=2$, $b=-3$, $c=1$ and $k--5$}
\end{minipage}
\end{figure}

\begin{figure}[H]
\centering
\begin{minipage}[b]{0.4\textwidth}
    \includegraphics[width=\textwidth]{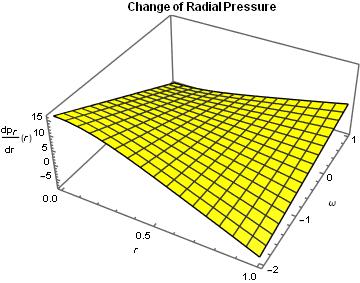}
    \caption{Graph for Change of Radial Pressure with $a=2$, $b=-3$, $c=1$ and $k--5$}
\end{minipage}
\hfill
\begin{minipage}[b]{0.4\textwidth}
    \includegraphics[width=\textwidth]{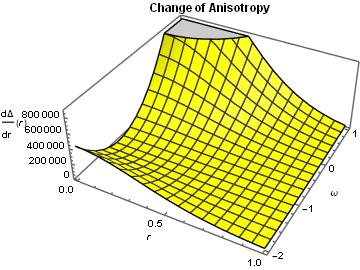}
    \caption{Graph for Change of Anisotropy with $a=2$, $b=-3$, $c=1$ and $k--5$}
\end{minipage}
\end{figure}

\subsection{Energy Conditions}
Raychaudhuri equations in cosmic fluid dynamics provides the
following energy conditions against the cosmic evolution.\cite{33}
\begin{center}
$R_{\mu\nu}u^{\mu}u^{\nu}\geq0$ and $R_{\mu\nu}n^{\mu}n^{\nu}\geq0$
\end{center}\par
This conditions can be again written as follows.
\begin{itemize}
    \item [1] Null energy condition (NEC) or $\rho+p\geq0$
    \item [2] Weak energy condition (WEC) or $\rho\geq0$ and $\rho+p\geq0$
    \item [3] Strong energy condition (SEC) or $\rho+3p\geq0$ and $\rho+p\geq0$
    \item [4] Dominant energy condition (DEC) or $\rho\geq0$ and $-\rho\leq p\leq\rho$
\end{itemize}

\begin{figure}[H]
\centering
\begin{minipage}[b]{0.4\textwidth}
    \includegraphics[width=\textwidth]{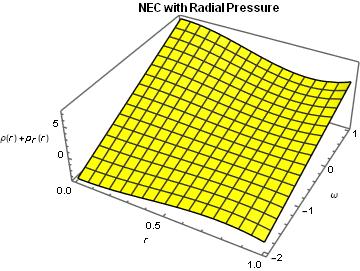}
    \caption{Graph for NEC with radial pressure with $a=2$, $b=-3$, $c=1$ and $k--5$}
\end{minipage}
\hfill
\begin{minipage}[b]{0.4\textwidth}
    \includegraphics[width=\textwidth]{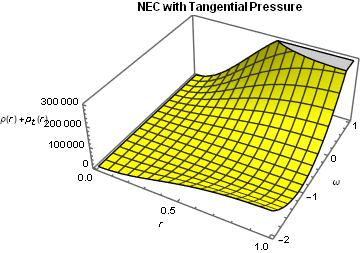}
    \caption{Graph for NEC with tangential pressure with $a=2$, $b=-3$, $c=1$ and $k--5$}
\end{minipage}
\end{figure}

\begin{figure}[H]
\centering
\begin{minipage}[b]{0.4\textwidth}
    \includegraphics[width=\textwidth]{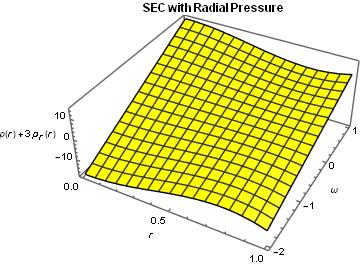}
    \caption{Graph for SEC with radial pressure with $a=2$, $b=-3$, $c=1$ and $k--5$}
\end{minipage}
\hfill
\begin{minipage}[b]{0.4\textwidth}
    \includegraphics[width=\textwidth]{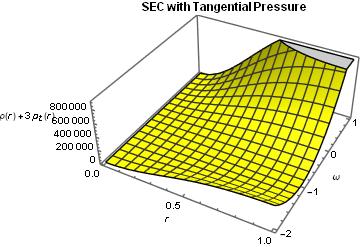}
    \caption{Graph for SEC with tangential pressure with $a=2$, $b=-3$, $c=1$ and $k--5$}
\end{minipage}
\end{figure}

\begin{figure}[H]
\centering
\begin{minipage}[b]{0.4\textwidth}
    \includegraphics[width=\textwidth]{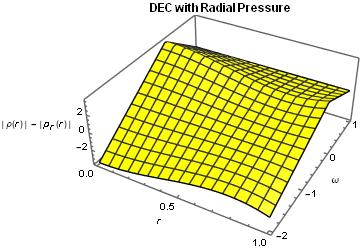}
    \caption{Graph for DEC with radial pressure with $a=2$, $b=-3$, $c=1$ and $k--5$}
\end{minipage}
\hfill
\begin{minipage}[b]{0.4\textwidth}
    \includegraphics[width=\textwidth]{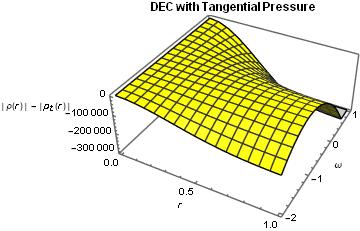}
    \caption{Graph for DEC with tangential pressure with $a=2$, $b=-3$, $c=1$ and $k--5$}
\end{minipage}
\end{figure}

\subsection{Stability Under Forces}
For a compact body, the stability of the system is important otherwise the volume or structure will become variable. Hence we need to make the resultant force into zero for the interior fluid system. There are mainly 4 types of forces should active here. They are as follows.
\begin{itemize}
    \item Hydrostatic Forces ($F_h=-\frac{dp_r}{dr}$)
    \item Gravitational Forces ($F_g=-\frac{\nu'}{2}(\rho+p_r)$)
    \item Anisotropic Repulsion Force ($F_a=\frac{2}{r}(p_t-p_r)$)
    \item Electric Force ($F_e=\frac{1}{4\pi}(\frac{2}{r}E^2+\frac{1}{2}\frac{d}{dr}(E^2))$)
\end{itemize}
Hence the Resultant force and the equilibrium condition can be written as follows with following the TOV (Tolman-Oppenheimer-Volkoff) equation.
\begin{equation}
    F_g+F_h+F_a+F_e=-\frac{\nu'}{2}(\rho+p_r)-\frac{dp_r}{dr}+\frac{2}{r}(p_t-p_r)+\frac{1}{4\pi}(\frac{2}{r}E^2+\frac{1}{2}\frac{d}{dr}(E^2))=0
\end{equation}

\begin{figure}[H]
\centering
\begin{minipage}[b]{0.8\textwidth}
    \includegraphics[width=\textwidth]{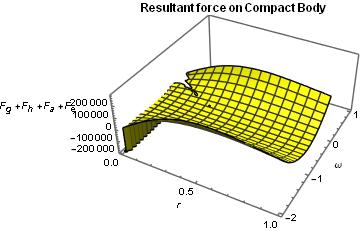}
    \caption{Graph for Resultant force according to TOV equation with $a=2$, $b=-3$, $c=1$ and $k--5$}
\end{minipage}
\end{figure}

\section{Interior thermodynamics analysis and radial dependencies of thermodynamic parameters of this compact body}
We have discussed the acceptability conditions and stability of our solutions in terms of geometric parameters. In general a physical system is ultimately stable when its thermodynamics make it the same. Hence we need to discuss the interior thermodynamics of the compact body of our work. We need to visualize the radial variations of the thermodynamics parameters to get the conditions of thermodynamics stability. Hence we mainly discuss the internal energy, temperature and entropy of the interior system. Moreover we search the conditions behind the equilibrium of this compact star. The thermodynamics should start from the relations between energy density and pressure with thermodynamical parameters. Hence we may get as follows.
\begin{equation}
    \rho=\frac{U}{V}
\end{equation}
And,
\begin{equation}
    p=-(\frac{\partial U}{\partial V})_s
\end{equation}
Here, $U=$ Internal energy and $V=$ volume of the system with the relation of $V=\frac{4}{3}\pi(\frac{x}{c})^{3/2}$. As we have two pressures i.e. radial and transverse, hence we can have two internal energies. If we consider the equation of state parameter as constant i.e. $\omega=$ constant we can find the internal energies as follows using the relations of pressure, density and anisotropy. Here we are considering $\frac{\Delta}{\rho}=f(x)$.
\begin{equation}
    U_t=U_{t0}\exp{(-\frac{3}{2}\int_{0}^{x}{\frac{f(x)}{x}dx}+\omega\int_{0}^{x}{\frac{1}{x}dx})}
\end{equation}
And the radial internal energy will be as follows.
\begin{equation}
    U_r=U_{r0}\exp{(-\frac{3}{2}\int_{0}^{x}{\frac{f(x)}{x}dx})}
\end{equation}
Now we can observe that for constant EOS parameter we can not get non-singular transverse internal energy as because $\ln{x}\rightarrow\infty$ with $x\rightarrow 0$. Thus, we we must have to choose the $\omega$ as a function of $x$. Hence let us consider the following relation.
\begin{equation}
    \omega=g(x)
\end{equation}
Hence, the transverse component can be written again as follows.
\begin{equation}
    U_t=U_{t0}\exp{(-\frac{3}{2}\int_{0}^{x}{\frac{(f(x)+g(x))}{x}dx})}
\end{equation}
Now the spatial dependent temperature can be defined with the relation as follows.
\begin{equation}
    \frac{dT}{T}=\frac{dm}{m}\frac{\partial p}{\partial \rho}
\end{equation}
Now here we got $m=$ the number of particles in the volume of compact bodies. Hence we can get $m\frac{4}{}\pi x^3=$ Number density at radius $x=$ constant where $x=$ radius of a shell inside of the compact body. Hence we have,
\begin{equation}
    \frac{dT}{T}=-\frac{dx}{x}\frac{\partial p}{\partial \rho}
\end{equation}
Hence the radial and transverse temperatures can be written as follows.
\begin{equation}
    T_r=T_{0r}\exp{(-\frac{9}{2}\int_{0}^{x}{\frac{1}{x}\frac{\partial p_r}{\partial \rho}dx})}
\end{equation}
And,
\begin{equation}
    T_t=T_{0t}\exp{(-\frac{9}{2}\int_{0}^{x}{\frac{1}{x}\frac{\partial p_t}{\partial \rho}dx})}
\end{equation}
Now finally we can find the spatial dependent entropy of the interior fluid system of that compact body. From the first law of generalized thermodynamics we know that $ds=\frac{dU}{T}+\frac{pdV}{T}$. Hence from this we can write the transverse and radial entropies of that system are as follows.
\begin{equation}
    s_r=\int_{0}^{x}{\frac{dU_r}{T_r}}+2\pi c^{-3/2}\int_{0}^{x}{\frac{p_r}{T_r}x^{1/2}dx}
\end{equation}
And,
\begin{equation}
    s_t=\int_{0}^{x}{\frac{dU_t}{T_t}}+2\pi c^{-3/2}\int_{0}^{x}{\frac{p_t}{T_t}x^{1/2}dx}
\end{equation}
Hence with the choice of equation of state parameter as a function of $r$ or $x$ we can find the functional forms of all the parameters like internal energies, temperatures and entropies. Let us choose the most simplest version of EOS parameter as follows.
\begin{equation}
    \omega(x)=g(x)=\lambda x
\end{equation}

\section{Scalar Field Theory Corresponding this New Compact Star Problem}
Now we establish the spatial dependent scalar field theory from this newly derived compact body. The whole interior fluid dynamics can be discussed with the space dependent scalar fields. From the usual action of scalar field theory we know that the energy density and pressure can be written as follows.
\begin{equation}
    \rho_{\phi}=\frac{1}{2}(\dot{\phi})^2+V(\phi)
\end{equation}
And,
\begin{equation}
    p_{\phi}=\frac{1}{2}(\dot{\phi})^2-V(\phi)
\end{equation}
Now substituting those energy density and pressure with radial and transverse energy density and pressure we can easily find the radial and transverse components of scalar field kinetic energy and potential which are as follows.
\begin{equation}
    \frac{1}{2}(\nabla\phi)^{2}_{r}=\frac{1}{2}(\frac{\partial\phi}{\partial r})^2=\rho+p_r
\end{equation}
And,
\begin{equation}
    \frac{1}{2}(\nabla\phi)^{2}_{t}=\frac{1}{2}(\frac{1}{r^2}(\frac{\partial\phi}{\partial\theta})^2+\frac{1}{r^2\sin^2{\theta}}(\frac{\partial\phi}{\partial\psi})^2)=\rho+p_t
\end{equation}
Here the coordinate $\psi$ is the transverse coordinate of spherical polar system. Now the potential can be written as follows.
\begin{equation}
    V(\phi)_r=\rho-p_r
\end{equation}
And,
\begin{equation}
    V(\phi)_t=\rho-p_t
\end{equation}
The temporal part of the scalar field Lagrangian has been avoided as because the system in only space dependent. The graphical representations can be provided as follows.

\begin{figure}[H]
\centering
\begin{minipage}[b]{0.4\textwidth}
    \includegraphics[width=\textwidth]{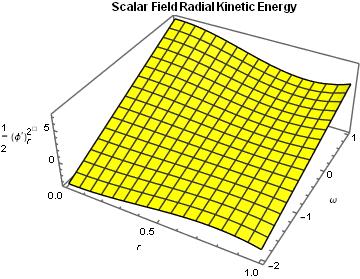}
    \caption{Graph for Scalar field Kinetic energy with radial pressure with $a=2$, $b=-3$, $c=1$ and $k--5$}
\end{minipage}
\hfill
\begin{minipage}[b]{0.4\textwidth}
    \includegraphics[width=\textwidth]{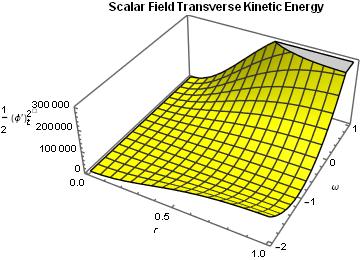}
    \caption{Graph for Scalar field kinetic energy with tangential pressure with $a=2$, $b=-3$, $c=1$ and $k--5$}
\end{minipage}
\end{figure}

\begin{figure}[H]
\centering
\begin{minipage}[b]{0.4\textwidth}
    \includegraphics[width=\textwidth]{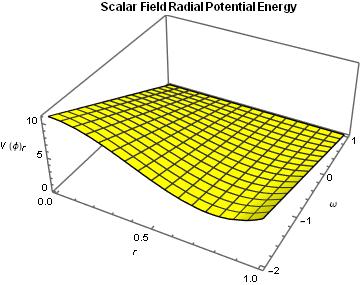}
    \caption{Graph for Scalar field potential energy with radial pressure with $a=2$, $b=-3$, $c=1$ and $k--5$}
\end{minipage}
\hfill
\begin{minipage}[b]{0.4\textwidth}
    \includegraphics[width=\textwidth]{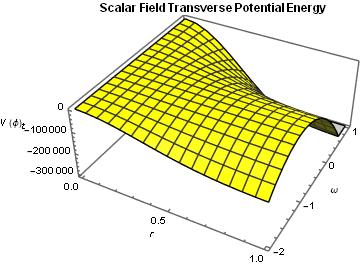}
    \caption{Graph for Scalar field potential energy with tangential pressure with $a=2$, $b=-3$, $c=1$ and $k--5$}
\end{minipage}
\end{figure}

\section{Interpretation of mass change with exterior Multi-fluid mass accretion and time dependency in scalar field theory}
Now we consider the mass accretion and its effect on the compact body stability. We consider expanding universe model to demonstrate the exterior fluid evolution that again insist in compact body mass accretion and its instability in structure. We can also provide the possible mass blast due to such instability. We know that the mass accretion is completely dependent upon the external cosmic fluids and their properties. Considering the universe contains the fluid with energy density and pressure as $\rho_{tot}$ and $p_{tot}$ we can introduce the mass accretion formalism as follows. Here the subscript "tot" is used to define the multi-fluid contribution on mass accretion.
\begin{equation}
    \dot{M}_1(t)=4\pi A(\rho_{tot}+p_{tot})
\end{equation}
Now we assume that the mass accretion will increase the total mass of the compact body and hence we can say that the mass accretion will change the radius of the compact system. This can be written as follows.
\begin{equation}
    M(R)=M_1(t)
\end{equation}
Hence we can write the time varying radius as follows.
\begin{equation}
    R(t)=M^{-1}(M_1(t))
\end{equation}
The time variation can be found from the solution of polynomial equation of $R$ found from equation 5. The radius is time varying, hence all other parameters and variables discussed in this work became time varying. Therefore we can find time varying scalar field model from time varying energy densities and pressures. The scalar field kinetic energy and potential energy can be written as follows.
\begin{equation}
    \frac{1}{2}(\dot{\phi})^2=\frac{1}{2}(\frac{1}{2}(\dot{\phi})^{2}_r+\frac{1}{2}(\dot{\phi})^{2}_t)=\rho(t)+\frac{1}{2}(p_r(t)+p_t(t))
\end{equation}
And,
\begin{equation}
    V(t)=V(\phi)=\frac{1}{2}(V_r(t)+V_t(t))=\rho(t)-\frac{1}{2}(p_r(t)+p_t(t))
\end{equation}
Hence we can observe that the scalar field theory for compact body is evolving due to continuous mass accretion. Even after the evolution the equilibrium conditions remain unaltered and this is because the change of mass mainly changes the radius which apparently changes everything with time. That means we can say that the mass accretion changes everything indirectly except the radius and for each instantaneous time the equilibrium conditions are keeping the model stable.

\section{Result Analysis and Discussion}
In section 2.3 we have discussed the path dependent variations of the astrophysical fluid variables $\rho$, $p_r$, $p_t$, $\omega$, $\Delta$ and $M$. We have shown their variations pictorially from figure 1 to 7. For our assumed parametric values we found positive values of energy density which is dissimilar to the exotic matter system in strange stars or dark energy stars. From the radial and transverse pressure graphical representations we can find positive pressures. Hence, our system can not be stabilized with the pressures from interior fluids and that is why we shall need high values of hydrodynamical forces for the same. The anisotropy remains positive throughout the whole region and thus we may conclude that the system stability may come with anisotropy also. The mass function and shearing scalar both provide positive values and thus validates of our model. In all those graphs (fig. 1 to 22) we have shown the values for different EOS parameter $\omega$. Basically we have used changing EOS parameter which changes only during phase transitions.\par
In the section 3.3 from figures 8 to 11 we can observe that our model violates the maximally and regularity conditions. The variables $\rho$, $p_r$, $p_t$ and $\Delta$ all of them increases with radial distances and hence they have higher values on surface that center. This may bring structure instabilities of the compact body discussed through our model. Although from figure 10, we can observe that for positive values of $\omega$, $\frac{dp_r}{dr}$ is less than zero around center and surface but providing $> 0$ in the middle shell volume.\par
We have discussed the energy conditions in section 3.4. We have found that both radial and transverse pressures are satisfying the strong, weak and null energy conditions with energy densities (fig. 12 to 17). But The dominant energy condition has been violated with tangential pressure. The satisfaction of energy conditions again proved that the fluid system discussed here can not provide repulsive behaviour in compact body.\par
In section 3.5 we have incorporated four forces i.e. Hydrostatic force, Gravitational force, Anisotropic repulsion force and Electric force to discuss the stability of the compact body system. We have to make it zero on the surface of the system. From figure 18 we can observe that all the forces are neutralized with the increase in radius and thus becoming stable system. The hydrostatic force and anisotropic force neutralizes all other attractive forces discussed here.\par
In section 4 we have discussed the interior thermodynamics through internal energy, temperature and entropy for both radial and tangential pressures. We have discussed the internal energy variation for both constant and variable EOS parameter. We have proved that the constant EOS parameter can not stabilize the compact body interior system and this is because the internal energy diverges for constant $\omega$. \par
In section 5 we have discussed the compact body interior fluid corresponding to scalar field theory. The scalar field and potential are different for different pressures. We have shown the variation of both tangential and radial kinetic and potential energies for changing EOS parameter. The Potential energies are seemed to be decreasing in nature with radius.

\section{Concluding Remarks}
The present work has been established with the solution of Einstein-Maxwell field equation using Buchdahl solution technique. We have considered nonlinear electromagnetic field whose nature is different  from Coulomb's inverse square law. The Buchdahl function assumption brought a new kind of transformations that also helped us to solve six variables solutions from six equations. The final results provide us a new kind of compact body where we have also tried to incorporate its inner fluid thermodynamics and stability analysis. Except the regularity conditions we have successfully satisfied all other conditions with our newly derived model. At the very end, we have introduced external cosmic fluid mass accretion of the compact body and proved that the mass accretion can increase the body radius of the system. The interior thermodynamics proved that a stable compact body can never contain constant equation of state parameter throughout its whole interior. 

\section*{Limitations of this work and Future Possibilities}
We have limited our work to the assumption of new type of Buchdahl Spacetime function that can produce a new type of compact star solution of Einstein-Maxwell equation. We have also studied the corresponding scalar field solution, There is an extensive application in future of this work to analyze several pulsars, Nebula Gamma Ray Blast and compact object analysis. All those areas are beyond the scope of this paper. The mass function and Electromagnetic Fields from such objects of our work can also be used in future to discuss the Gamma Ray Blast Shock wave, Supernovae Shock waves and radiations from them. In future we can also use the mass function to interpret the cosmic evolution and discuss the mass accretion of that compact system with universe evolution.


\begin{thebibliography}{82}


\bibitem{1}Rej, P. and Bhar, P., 2021. Charged strange star in f (R, T) $ f (R, T) $ gravity with linear equation of state. Astrophysics and Space Science, 366(4), pp.1-17.
\bibitem{2}Rej, P., Bhar, P. and Govender, M., 2021. Charged compact star in f (R, T) gravity in Tolman–Kuchowicz spacetime. The European Physical Journal C, 81(4), pp.1-15.
\bibitem{3}Bhar, P., Murad, M.H. and Pant, N., 2015. Relativistic anisotropic stellar models with Tolman VII spacetime. Astrophysics and Space Science, 359(1), pp.1-9.




\bibitem{4}Bowers, R.L. and Liang, E.P.T., 1974. Anisotropic spheres in general relativity. The Astrophysical Journal, 188, p.657.
\bibitem{5}Gupta, Y.K. and Maurya, S.K., 2011. A class of charged analogues of Durgapal and Fuloria superdense star. Astrophysics and Space Science, 331(1), pp.135-144.
\bibitem{6}Ivanov, B.V., 2021. Generating solutions for charged stellar models in general relativity. The European Physical Journal C, 81(3), pp.1-8.
\bibitem{7}Herrera, L., Ruggeri, G.J. and Witten, L., 1979. Adiabatic contraction of anisotropic spheres in general relativity. The Astrophysical Journal, 234, pp.1094-1099.
\bibitem{8}Kiess, T.E., 2012. Exact physical Maxwell-Einstein Tolman-VII solution and its use in stellar models. Astrophysics and Space Science, 339(2), pp.329-338.
\bibitem{9}Herrera, L. and Nunez, L., 1989. Modeling'hydrodynamic phase transitions' in a radiating spherically symmetric distribution of matter. The Astrophysical Journal, 339, pp.339-353.
\bibitem{10}Maurya, S.K., Maharaj, S.D., Kumar, J. and Prasad, A.K., 2019. Effect of pressure anisotropy on Buchdahl-type relativistic compact stars. General Relativity and Gravitation, 51(7), pp.1-28.
\bibitem{11}Sharma, R., Ghosh, A., Bhattacharya, S. and Das, S., 2021. Anisotropic generalization of Buchdahl bound for specific stellar models. The European Physical Journal C, 81(6), pp.1-5.
\bibitem{12}Prasad, A.K. and Kumar, J., 2021. Anisotropic relativistic fluid spheres in the Buchdahl model. arXiv preprint arXiv:2103.12583.
\bibitem{13}Prasad, A.K., Kumar, J. and Sarkar, A., 2021. Behavior of anisotropic fluids with Chaplygin equation of state in Buchdahl spacetime. arXiv preprint arXiv:2104.13004.
\bibitem{14}Tolman, R.C., 1939. Static solutions of Einstein's field equations for spheres of fluid. Physical Review, 55(4), p.364.
\bibitem{15}Esculpi, M., Malaver, M. and Aloma, E., 2007. A comparative analysis of the adiabatic stability of anisotropic spherically symmetric solutions in general relativity. General Relativity and Gravitation, 39(5), pp.633-652.
\bibitem{16}Feroze, T. and Siddiqui, A.A., 2011. Charged anisotropic matter with quadratic equation of state. General Relativity and Gravitation, 43(4), pp.1025-1035.
\bibitem{17}Oppenheimer, J.R. and Volkoff, G.M., 1939. On massive neutron cores. Physical Review, 55(4), p.374.
\bibitem{18}Buchdahl, H.A., 1959. General relativistic fluid spheres. Physical Review, 116(4), p.1027.
\bibitem{19}Cosenza, M., Herrera, L., Esculpi, M. and Witten, L., 1982. Evolution of radiating anisotropic spheres in general relativity. Physical Review D, 25(10), p.2527.
\bibitem{20}Durgapal, M.C. and Bannerji, R., 1983. New analytical stellar model in general relativity. Physical Review D, 27(2), p.328.
\bibitem{21}Komathiraj, K. and Maharaj, S.D., 2007. Analytical models for quark stars. International Journal of Modern Physics D, 16(11), pp.1803-1811.
\bibitem{22}Sunzu, J.M. and Danford, P., 2017. New exact models for anisotropic matter with electric field. Pramana, 89(3), pp.1-9.
\bibitem{23}Sunzu, J.M., Maharaj, S.D. and Ray, S., 2014. Quark star model with charged anisotropic matter. Astrophysics and Space Science, 354(2), pp.517-524.
\bibitem{24}Ivanov, B.V., 2002. Static charged perfect fluid spheres in general relativity. Physical Review D, 65(10), p.104001.
\bibitem{25}Takisa, P.M. and Maharaj, S.D., 2013. Some charged polytropic models. General Relativity and Gravitation, 45(10), pp.1951-1969.
\bibitem{26}Tello-Ortiz, F., Malaver, M., Rincón, Á. and Gomez-Leyton, Y., 2020. Relativistic anisotropic fluid spheres satisfying a non-linear equation of state. The European Physical Journal C, 80(5), pp.1-13.
\bibitem{27}Maurya, S.K., Banerjee, A., Jasim, M.K., Kumar, J., Prasad, A.K. and Pradhan, A., 2019. Anisotropic compact stars in the Buchdahl model: A comprehensive study. Physical Review D, 99(4), p.044029.
\bibitem{28}Prasad, A.K., Kumar, J. and Sarkar, A., 2021. Behavior of anisotropic fluids with Chaplygin equation of state in Buchdahl spacetime. arXiv preprint arXiv:2104.13004.
\bibitem{29}Thirukkanesh, S. and Maharaj, S.D., 2008. Charged anisotropic matter with a linear equation of state. Classical and Quantum Gravity, 25(23), p.235001.
\bibitem{30}Thirukkanesh, S. and Ragel, F.C., 2012. Exact anisotropic sphere with polytropic equation of state. Pramana, 78(5), pp.687-696.





\bibitem{31}Sadhukhan. S, Quintessence Model Calculations for Bulk Viscous Fluid and Low Value Predictions of the Coefficient of Bulk Viscosity, International Journal of Science and Research (IJSR) 9(3):1419-1420, DOI: 10.21275/SR20327132301
\bibitem{32}A. Kar, S. Sadhukhan, Hamiltonian Formalism for Bianchi Type
I Model for Perfect Fluid as Well as for the Fluid with Bulk and
Shearing Viscosity. Basic and Applied Sciences into Next Frontiers
(New Delhi Publishers, 2021) (ISBN: 978-81-948993-0-3)
\bibitem{33}A. Kar, S. Sadhukhan, Quintessence model with bulk viscosity and
some predictions on the coefficient of bulk viscosity and gravitational constant, recent advancement of mathematics in science and
technology (2021) (ISBN: 978-81-950475-0-5)
\bibitem{34}Kar, A., Sadhukhan, S. and Chattopadhyay, S., 2021. Energy conditions for inhomogeneous EOS and its thermodynamics analysis with the resolution on finite time future singularity problems. International Journal of Geometric Methods in Modern Physics, p.2150131.
\bibitem{35}Kar, A., Sadhukhan, S. and Chattopadhyay, S., 2021. Thermodynamics and energy condition analysis for Van-Der-Waals EOS without viscous cosmology. Physica Scripta. https://doi.org/10.1088/1402-4896/ac2f00
\bibitem{36}Kar, A., Sadhukhan, S. and Debnath, U., 2021. Condensed body mass accretion with DBI-essence dark energy and its reconstruction with f (Q) gravity. arXiv preprint arXiv:2109.10906.
\bibitem{37}Sadhukhan, S., Kar, A. and Chattopadhay, S., 2021. Thermodynamic analysis for Non-linear system (Van-der-Waals EOS) with viscous cosmology. The European Physical Journal C, 81(10), pp.1-21. arXiv:2110.13831
\bibitem{38}Kar, A., Sadhukhan, S. and Debnath, U., 2021. Reconstruction of DBI-essence dark energy with f ( R ) gravity and its effect on black hole and wormhole mass accretion. Modern Physics Letters A, Vol. 36, No. 38 (2021) 2150262 (17 pages) ; DOI: 10.1142/S021773232150262X
\bibitem{39}Malaver, M., Kasmaei, H.D., Iyer, R., Sadhukhan, S. and Kar, A., 2021. A theoretical model of Dark Energy Stars in Einstein-Gauss-Bonnet Gravity. arXiv preprint arXiv:2106.09520.
\bibitem{40}Kar, A. and Gupta, N., 2021. Ultra-high Energy Gamma-rays from Past Explosions in our Galaxy. arXiv preprint arXiv:2112.08757.



\bibitem{41}Kumar, J. and Bharti, P., 2021. The classification of interior anisotropic fluid solutions. arXiv preprint arXiv:2112.12518.




\bibitem{42}Malaver, M. Black Holes, Wormholes and Dark Energy Stars in General Relativity. Lambert Academic Publishing, Berlin. ISBN: 978-3-659-34784-9, 2013
\bibitem{43}Malaver, M.; Kasmaei, H.D. Relativistic stellar models with quadratic equation of state. International Journal of Mathematical Modelling and Computations. 2020, 10, 111-124.
\bibitem{44}Malaver, M. New Mathematical Models of Compact Stars with Charge Distributions. International Journal of Systems Science and Applied Mathematics. 2017, 2, 93-98, DOI: 10.11648/j.ijssam.20170205.13.
\bibitem{45}Malaver, M. Generalized Nonsingular Model for Compact Stars Electrically Charged. World Scientific News. 2018, 92, 327-339.
\bibitem{46}Malaver, M. Analytical models for compact stars with a linear equation of state. World Scientific News, 2016, 50, 64-73.
\bibitem{47}Malaver, M. Analytical model for charged polytropic stars with Van der Waals Modified Equation of State. American Journal of Astronomy and Astrophysics. 2013, 1, 37-42.
\bibitem{48}Malaver, M. Quark Star Model with Charge Distributions. Open Science Journal of Modern Physics. 2014, 1, 6-11.
\bibitem{49}Malaver, M. Strange Quark Star Model with Quadratic Equation of State. Frontiers of Mathematics and Its Applications. 2014, 1, 9-15.
\bibitem{50}Malaver, M. Charged anisotropic models in a modified Tolman IV space time. World Scientific News. 2018, 101, 31-43
\bibitem{51}Malaver, M. Charged stellar model with a prescribed form of metric function y(x) in a Tolman VII spacetime. World Scientific News.2018, 108, 41-52
\bibitem{52}Malaver, M. Classes of relativistic stars with quadratic equation of state. World Scientific News. 2016, 57, 70 -80
\bibitem{53}Malaver, M. Some new models of anisotropic compact stars with quadratic equation of state. World Scientific News. 2018, 109, 180-194
\bibitem{54}Malaver, M. Charged anisotropic matter with modified Tolman IV potential. Open Science Journal of Modern Physics. 2015, 2(5), 65-71


\bibitem{55}Abbott, B.P., Abbott, R., Abbott, T.D., Abraham, S., Acernese, F., Ackley, K., Adams, C., Adhikari, R.X., Adya, V.B., Affeldt, C. and Agathos, M., 2020. GW190425: Observation of a compact binary coalescence with total mass$\sim 3.4 M\odot$ The Astrophysical Journal Letters, 892(1), p.L3.
\bibitem{56}Bailyn, C.D., Jain, R.K., Coppi, P. and Orosz, J.A., 1998. The mass distribution of stellar black holes. The Astrophysical Journal, 499(1), p.367.
\bibitem{57}Özel, F., Psaltis, D., Narayan, R. and McClintock, J.E., 2010. The black hole mass distribution in the galaxy. The Astrophysical Journal, 725(2), p.1918.
\bibitem{58}Belczynski, K., Wiktorowicz, G., Fryer, C.L., Holz, D.E. and Kalogera, V., 2012. Missing black holes unveil the supernova explosion mechanism. The Astrophysical Journal, 757(1), p.91.
\bibitem{59}Özel, F., Baym, G. and Güver, T., 2010. Astrophysical measurement of the equation of state of neutron star matter. Physical Review D, 82(10), p.101301.
\bibitem{60}Farr, W.M., Sravan, N., Cantrell, A., Kreidberg, L., Bailyn, C.D., Mandel, I. and Kalogera, V., 2011. The mass distribution of stellar-mass black holes. The Astrophysical Journal, 741(2), p.103.
\bibitem{61}Müller, H. and Serot, B.D., 1996. Relativistic mean-field theory and the high-density nuclear equation of stateNucl. Phys. A606, 508.
\bibitem{62}Rhoades Jr, C.E. and Ruffini, R., 1974. Maximum mass of a neutron star. Physical Review Letters, 32(6), p.324.
\bibitem{63}Özel, F., Psaltis, D., Narayan, R. and Villarreal, A.S., 2012. On the mass distribution and birth masses of neutron stars. The Astrophysical Journal, 757(1), p.55.
\bibitem{64}Kiziltan, B., Kottas, A., De Yoreo, M. and Thorsett, S.E., 2013. The neutron star mass distribution. The Astrophysical Journal, 778(1), p.66.
\bibitem{65}Abbott, B.P., Abbott, R., Abbott, T.D., Acernese, F., Ackley, K., Adams, C., Adams, T., Addesso, P., Adhikari, R.X., Adya, V.B. and Affeldt, C., 2018. GW170817: Measurements of neutron star radii and equation of state. Physical review letters, 121(16), p.161101.
\bibitem{66}Abbott, B.P., Abbott, R., Abbott, T.D., Acernese, F., Ackley, K., Adams, C., Adams, T., Addesso, P., Adhikari, R.X., Adya, V.B. and Affeldt, C., 2019. Tests of general relativity with GW170817. Physical review letters, 123(1), p.011102.
\bibitem{67}Antoniadis, J., Freire, P.C., Wex, N., Tauris, T.M., Lynch, R.S., Van Kerkwijk, M.H., Kramer, M., Bassa, C., Dhillon, V.S., Driebe, T. and Hessels, J.W., 2013. A massive pulsar in a compact relativistic binary. Science, 340(6131).
\bibitem{68}Ruderman, M., 1972. Pulsars: structure and dynamics. Annual Review of Astronomy and Astrophysics, 10(1), pp.427-476.
\bibitem{69}Sokolov, A.I., 1980. Phase transitions in a superfluid neutron liquid. Sov. Phys. JETP, 52(4), p.575.
\bibitem{70}Sawyer, R.F., 1972. Condensed $\pi-$ phase in neutron-star matter. Physical Review Letters, 29(6), p.382.
\bibitem{71}Abeysekara, A.U., Albert, A., Alfaro, R., Camacho, J.A., Arteaga-Velázquez, J.C., Arunbabu, K.P., Rojas, D.A., Solares, H.A., Baghmanyan, V., Belmont-Moreno, E. and BenZvi, S.Y., 2020. Multiple galactic sources with emission above 56 TeV detected by HAWC. Physical review letters, 124(2), p.021102.
\bibitem{72}Mondal, S.K., Prince, R., Gupta, N. and Das, A.K., 2021. Spectral Modeling of Flares in Long-term Gamma-Ray Light Curve of PKS 0903-57. The Astrophysical Journal, 922(2), p.160.
\bibitem{73}Baring, M.G., Ellison, D.C., Reynolds, S.P., Grenier, I.A. and Goret, P., 1999. Radio to gamma-ray emission from shell-type supernova remnants: predictions from nonlinear shock acceleration models. The Astrophysical Journal, 513(1), p.311.
\bibitem{74}Blumenthal, G.R. and Gould, R.J., 1970. Bremsstrahlung, synchrotron radiation, and compton scattering of high-energy electrons traversing dilute gases. Reviews of Modern Physics, 42(2), p.237.
\bibitem{75}Breuhaus, M., Hahn, J., Romoli, C., et al. 2021, The
Astrophysical Journal Letters, 908, L49
\bibitem{76}LHAASO collaboration, 2021. Discovery of the Ultra-high energy gamma-ray source LHAASO J2108+ 5157. arXiv preprint arXiv:2106.09865.
\bibitem{77}Cao, Z., Aharonian, F.A., An, Q., Bai, L.X., Bai, Y.X., Bao, Y.W., Bastieri, D., Bi, X.J., Bi, Y.J., Cai, H. and Cai, J.T., 2021. Ultrahigh-energy photons up to 1.4 petaelectronvolts from 12 Gamma ray Galactic sources. Nature, 594(7861), pp.33-36.
\bibitem{78}Hahn, J., 2016. GAMERA–a new modeling package for non-thermal spectral modeling, in proceedings of the 34th International Cosmic Ray Conference. PoS (ICRC2015), 917.
\bibitem{79}Cao, Z., Aharonian, F., An, Q., Bai, L.X., Bai, Y.X., Bao, Y.W., Bastieri, D., Bi, X.J., Bi, Y.J., Cai, H. and Cai, J.T., 2021. Discovery of a New Gamma-Ray Source, LHAASO J0341+ 5258, with Emission up to 200 TeV. The Astrophysical Journal Letters, 917(1), p.L4.
\bibitem{80}Tomar, G., Gupta, N. and Prince, R., 2021. Broadband Modeling of Low-luminosity Active Galactic Nuclei Detected in Gamma Rays. The Astrophysical Journal, 919(2), p.137.







\end{thebibliography}
\end{document}